\documentclass{article}
\usepackage{slashed,feynmf,epsfig,amsmath,amssymb,enumitem}
\usepackage[papersize={8.5in,11in}]{geometry}
\usepackage{graphicx}
\newcommand{\be}{\begin{equation}}
\newcommand{\ee}{\end{equation}}
\begin{document}
\title{Modified Gravity (MOG) and Heavy Neutron Star in Mass Gap}
\author{J. W. Moffat\\~\\
Perimeter Institute for Theoretical Physics, Waterloo, Ontario N2L 2Y5, Canada\\
and\\
Department of Physics and Astronomy, University of Waterloo, Waterloo,\\
Ontario N2L 3G1, Canada}
\maketitle




\begin{abstract}
The modified gravity (MOG) theory is applied to the gravitational wave binary merger GW190814 to demonstrate that the modified Tolman-Oppenheimer-Volkoff equation for a neutron star can produce a mass $M=2.6 -2.7 M_\odot$, allowing for the binary secondary component to be identified as a heavy neutron star in the hypothesized mass gap $2.5 - 5 M_\odot$.
\end{abstract}
\maketitle

\section{Introduction}

The present work applies a modified gravity (MOG), also called Scalar--Tensor--Vector--Gravity (STVG)~\cite{Moffat2006,Moffat2020} to the merging binary GW190814. The modified Tolman--Oppenheimer--Volkoff equation has been demonstrated to allow heavier neutron stars than General Relativity (GR)~\cite{Armengol2017,PerezRomero2020}. The MOG is described by a fully covariant action and field equations, extending GR by the addition of two gravitational degrees of freedom. The first is $G= 1/\chi$, where $G$ is the coupling strength of gravity and $\chi$ is a scalar field. The second degree of freedom is a massive gravitational vector field $\phi_\mu$. The gravitational coupling of the vector graviton to matter is universal with the gravitational charge $Q_g=\sqrt{\alpha G_N}M$, where $\alpha$ is a dimensionless scalar field, $G_N$ is Newton's gravitational constant and $M$ is the mass of a body.

The mass ratio of 0.112 of GW190814 makes it differ significantly from other compact binary mergers~\cite{Abbott2020}. It has no detected electromagnetic optical counterpart. The GW190814 compact companion mass lies in the lower mass gap of $2.5 - 5 M_\odot$~\cite{Ozel2012} between known neutron stars and black holes. It is more massive than the heaviest pulsar in the Galaxy with a mass $2.14^{+0.20}_{-0.18} M_\odot$~\cite{Cromartie2019} and more massive than the $M=1.61 - 2.52 M_\odot$ mass of the primary component of GW190425, an outlier of the observed Galactic population of binary neutron stars. Even small increases in neutron star masses are important for the fundamental understanding of their interiors and their equations of state (EOSs)~\cite{Ozel2016}. The mass $2.7 M_\odot$ of the black hole candidate merger remnant GW170817 is comparable to the mass of the secondary component of GW190814. Because of the uncertainties in the estimations of neutron star masses obtained by studies of nuclear and non-nuclear EOSs, it is generally thought that GW190814 is not a black hole neutron star merger. We will demonstrate that the MOG combined with known neutron star EOSs can produce heavier neutron stars and pulsars and resolve the GW190814 neutron star-black hole puzzle.

\section{The MOG Field Equations}

The MOG field equations will take a simplified form assuming that $G$ is a slowly varying constant and we will neglect for the compact neutron stars the small mass $\mu$~\cite{Moffat2006,Moffat2020}. The field equations are given by (we use the metric signature $(+,-,-,-)$ and units with $c=1$):
\be
\label{Gequation}
G_{\mu\nu}=8\pi GT_{\mu\nu},
\ee
\be
\label{Bequation}
\nabla_\nu B^{\mu\nu}=\frac{1}{\sqrt{-g}}\partial_\nu(\sqrt{-g}B^{\mu\nu})=J_M^\mu,
\ee
where $G_{\mu\nu}=R_{\mu\nu}-\frac{1}{2}R$, $G=G_N(1+\alpha)$, $\nabla_\mu$ denotes the covariant derivative with respect to the metric $g_{\mu\nu}$, 
$g={\rm det}(g_{\mu\nu})$ and $B_{\mu\nu}=\partial_\mu\phi_\nu-\partial_\nu\phi_\mu$.  Moreover, $J_M^\mu=\sqrt{\alpha G_N}\rho u^\mu$, $\rho$ is the density of matter and field energy and $u^\mu=dx^\mu/ds$. The energy-momentum tensor is
\be
T_{\mu\nu}=T^M_{\mu\nu}+T^\phi_{\mu\nu},
\ee
where
\be
T^\phi_{\mu\nu}=-\biggl({B_\mu}^\alpha B_{\alpha\nu}-\frac{1}{4}g_{\mu\nu}B^{\alpha\beta}B_{\alpha\beta}\biggr).
\ee

The conservation equations are given by
\be
\label{conservequation}
\nabla_\nu(T^{\mu\nu}_M+T^{\mu\nu}_\phi)=0.
\ee

The gravitational coupling strength $G=G_N(1+\alpha)$ in the field equations (\ref{Gequation}) is chosen to be constant and scales with $\alpha$, depending on the strength of the gravitational field. This follows the procedure in ref.~\cite{Armengol2017} in which they determine the properties of compact neutron stars with constant values of the parameter $\alpha$. Moreover, for the following application to compact neutron stars, we ignore the small mass $\mu$, which corresponds to $m_\phi\sim 10^{-28}$ eV according to fits of MOG to galaxy rotation curves without dark matter~\cite{MoffatRahvar2013,MoffatRahvar2014,GreenMoffat2019,DavariRahvar2020,BrownsteinMoffat2007,IsraelMoffat2018,MoffatHaghighi2017}.

The modified Newtonian acceleration law for weak gravitational fields and for a point particle can be written
as~\cite{Moffat2006}:
\begin{equation}
\label{MOGaccelerationlaw}
a_{\rm MOG}(r)=-\frac{G_NM}{r^2}[1+\alpha-\alpha\exp(-\mu r)(1+\mu r)].
\end{equation}
This reduces to Newton's gravitational acceleration in the limit
$\mu r\ll 1$. In the limit that $r\rightarrow\infty$, we get from
(\ref{MOGaccelerationlaw}) for approximately constant $\alpha$ and
$\mu$:
\begin{equation}
\label{AsymptoticMOG}
a_{\rm MOG}(r)\approx -\frac{G_N(1+\alpha)M}{r^2}.
\end{equation}

\section{The Static Spherically Symmetric Matter Solution}

We consider a spacetime with a static, spherically symmetric manifold and model the stellar matter with a perfect fluid energy-momentum tensor. The modified Tolman--Oppenheimer--Volkoff equation is used to describe a neutron star using equations of state~\cite{Armengol2017}. The spacetime metric is described by
\be
ds^2=\exp(\nu(r))dt^2-\exp(\lambda(r))dr^2-r^2(d\theta^2+\sin^2\theta d\phi^2).
\ee
The stellar matter is modeled with a perfect fluid energy-momentum tensor:
\be
T^M_{\mu\nu}=(\rho(r)+p(r))u_\mu u_\nu-p(r)g_{\mu\nu},
\ee
where $\rho(r)$ and $p(r)$ denote the density and pressure of the fluid, respectively, and $u^\mu=(\exp(-\nu/2),0,0,0)$ denotes the four-velocity of a mass element. The four-current density is:
\be
J_M^\mu=4\pi\sqrt{\alpha G_N}\rho u^\mu=\frac{4\pi\sqrt{\alpha G_N}\rho}{\sqrt{g_{00}}}\frac{dx^\mu}{dt}.
\ee

The non-vanishing components of the vector field strength $B^{\mu\nu}$ are given by\footnote{The constant $\omega$ included in the field equations of STVG~\cite{Moffat2006} has been set equal to $1/\sqrt{12}$ and is omitted here.}
\be
B^{01}(r)=-B^{10}(r)=-\exp\biggl(-\frac{\nu(r)+\lambda(r)}{2}\biggr)\frac{Q_g(r)}{r^2},
\ee
\be
Q_g(r)=4\pi\int drr^2\exp(\lambda(r)/2)\sqrt{\alpha G_N}\rho(r).
\ee
The components of the vector field $\phi_\mu$ energy-momentum tensor are given by
\be
{T_\phi^0}_0={T_\phi^1}_1=-{T_\phi^2}_2=-{T_\phi^3}_3=\frac{1}{2}\frac{Q_g^2(r)}{r^4}.
\ee

Solving for the field equations (\ref{Gequation}) gives~\cite{Armengol2017}:
\be
\label{lambdaeq}
\exp(-\lambda(r))=1-\frac{2GM(r)}{c^2r}-\frac{1}{r}\frac{4\pi G}{c^4}\int dr\frac{Q_g(r)}{r^2},
\ee
\be
\label{nueq}
\nu(r)=-\lambda(r)+\frac{8\pi G}{c^4}\int dr\exp(\lambda(r)r(c^2\rho(r)+p(r)),
\ee
where we have reinstated the speed of light $c$ and as before $G=G_N(1+\alpha)$ and $Q_g=\sqrt{\alpha G_N}M$. For a point mass, we obtain the static spherically symmetric matter-free, $T_M^{\mu\nu}=J_M^\mu=0$, solution~\cite{Moffat2015}:
\be
ds^2=\biggl(1-\frac{2GM}{c^2r}-\frac{\alpha GG_NM^2}{c^4r^2}\biggr)dt^2-\biggl(1-\frac{2GM}{c^2r}-\frac{\alpha GG_NM^2}{c^4r^2}\biggr)^{-1}dr^2
-r^2(d\theta^2+\sin^2\theta d\phi^2).
\ee
By using the solutions (\ref{lambdaeq}) and (\ref{nueq}) and the conservation equation (\ref{conservequation}), the modified Tolman--Oppenheimer--Volkoff equation now takes the form~\cite{Armengol2017}:
\be
\label{TOVequation}
\frac{dP(r)}{dr}=-\frac{\exp(\lambda(r))}{r^2}\biggl(\frac{4\pi G}{c^4}p(r)r^3-\frac{2GQ_g^2(r)}{c^4r}+\frac{G M(r)}{c^2}
+\frac{2\pi G}{c^4}\int dr\frac{Q_g^2(r)}{r^2}\biggr)\\
(\rho(r)c^2+p(r))+\frac{Q_g(r)}{r^4}\frac{dQ_g(r)}{dr}.
\ee

\section{Equations of State and Neutron Stars}

The MOG Tolman--Oppenheimer--Volkoff equation (\ref{TOVequation}) is integrated numerically using known models of EOS, which relate the pressure $P(r)$ with the density $\rho(r)$ of each r shell. Four EOSs are investigated in ref.~\cite{Armengol2017} and the integration is carried out from the neutron star center up to its surface. The central densities are assumed to be in the range
$14.6 < \log(\rho_c[{\rm g cm^{-3}}]) < 15.9$. The deviations from GR are for non-zero values of constant $\alpha$, which can deviate in value from one shell to another depending on the mass of each shell.  In order to maintain astronomically acceptable masses and radii of neutron stars, $\alpha\lesssim 10^{-2}$. where $\rho(r)$ is determined by numerical integration of the Tolman--Oppenheimer-Volkoff equation with results given in ref.~\cite{Armengol2017}. When $\alpha=0$ the MOG solution reduces to the GR solution. In the modified Tolman--Oppenheimer--Volkoff equation gravitational attraction produced by negative terms in the derivative of the pressure, $dP(r)/dr$, is counteracted by positive $Q_g$ terms due the repulsive gravity.

The result is obtained that MOG predictions for the masses of neutron stars coincide with GR predictions for the high density neutron stars, reflecting the fact that MOG retrieves the classical results from the interchange of attraction and repulsion in the modified Tolman--Oppenheimer--Volkoff equation. When a certain maximum mass is reached, the gravitational attraction dominates and the density profile abruptly decreases. When the central density decreases differences arise from the predictions of GR that find its maximum mass at $\rho\sim 10^{15.3} {\rm g cm^{-3}}$. For lower values of density the predictions of MOG converge to those of GR. The results obtained show that for $\alpha\lesssim 10^{-2}$ maximum masses occur for which gravitational attraction dominates and produces masses exceeding those predicted by GR.

Importantly, the predicted MOG neutron star masses exceed GR masses for lower central densities than GR, although the radii predicted are ;larger than those predicted by GR. This means that solutions to the modified Tolman--Oppenheimer--Volkoff equation exist that can correspond to a heavy neutron star mass $M\sim 2.6-2.7 M_\odot$. This bigger mass can be identified with the compact object in the neutron star merger GW190814. Thus, MOG can interpret the companion compact object as a heavy neutron star as opposed to the object being a light black hole.

\section{Conclusions}

The hypothesized lower mass gap $2.5 - 2.7 M_\odot$ has been a puzzle for decades as it lies between the heaviest know neutron star and the lightest known black hole. It is possible that the mass gap may not exist but may be due to limitations in astronomer's observational capabilities. The GW190814 binary coalescence in a galaxy about 800 million light years away from earth, and the detection of gravitational waves emitted by the coalescence, fills this gap by the secondary compact object with a mass $2.6 -2.7 M_\odot$. Because of the maximum mass limitations of known nuclear EOSs and the GR relativistic Tolmann--Oppenheimer--Volkoff equation, it is generally accepted that the secondary companion is a very light black hole. The modified gravity MOG Tolman--Oppenheimer--Volkoff equation for a neutron star combined with known EOSs can produce more massive neutron stars than GR, depending on the MOG deviation parameter $\alpha\sim 10^{-2}$. For neutron star densities of medium size the modified Tolman--Oppenheimer--Volkoff equation can produce a neutron star mass $M\sim 2.7 M_\odot$ in the mass gap and resolve the mass gap conundrum by identifying the secondary object as a heavy neutron star.

\section*{Acknowledgments}

I thank Martin Green and Viktor Toth for helpful discussions. Research at the Perimeter Institute for Theoretical Physics is supported by the Government of Canada through industry Canada and by the Province of Ontario through the Ministry of Research and Innovation (MRI).

\end{document}